\def\year{2019}\relax
\documentclass[letterpaper]{article} 
\usepackage{aaai19}  
\usepackage{times}  
\usepackage{helvet}  
\usepackage{courier}  
\usepackage{url}  
\usepackage{graphicx}  

\usepackage{amssymb}
\usepackage{amsmath}
\usepackage{tikz}
\usetikzlibrary{calc}
\usepackage{algorithm}
\usepackage[noend]{algpseudocode}
\usepackage{thm-restate}
\usepackage{enumitem}
\usepackage{todonotes}
\usepackage{dsfont}

\newtheorem{theorem}{Theorem}
\newtheorem{corollary}{Corollary}[theorem]
\newtheorem{lemma}[theorem]{Lemma}
\newtheorem{definition}{Definition}

\DeclareMathOperator*{\argmax}{arg\,max}

\newcommand{\tr}{\mathsf{T}}

\begin{document}

\title{Computing Optimal Coarse Correlated Equilibria in Sequential Games}
\author{Andrea Celli\textsuperscript{\textnormal 1}, Stefano Coniglio\textsuperscript{\textnormal 2}, Nicola Gatti\textsuperscript{\textnormal 1}\\
\textsuperscript{1} Politecnico di Milano, Piazza Leonardo da Vinci 32, Milan, Italy \\
\textsuperscript{2}  University of Southampton, University Road SO17 1BJ, Southampton, United Kingdom \\
\{andrea.celli,nicola.gatti\}@polimi.it, s.coniglio@soton.ac.uk
}

\maketitle

\begin{abstract}
We investigate the computation of equilibria in extensive-form games where \emph{ex ante} correlation is possible, focusing on correlated equilibria requiring the least amount of communication between the players and the mediator.
Motivated by the hardness results on the computation of normal-form correlated equilibria, we introduce the notion of normal-form \emph{coarse} correlated equilibrium, extending the definition of coarse correlated equilibrium to sequential games.
We show that, in two-player games without chance moves, an optimal (e.g., social welfare maximizing) normal-form coarse correlated equilibrium can be computed in polynomial time, and that  in general multi-player games (including two-player games with Chance), the problem is \textsf{NP}-hard. 
For the former case, we provide a polynomial-time algorithm based on the ellipsoid method and also propose a more practical one, which can be efficiently applied to problems of considerable size. 
Then, we discuss how our algorithm can be extended to games with Chance and games with more than two players.
\end{abstract}

\section{Introduction}
The computational study of adversarial interactions is a central problem in Artificial Intelligence, aiming at finding players' optimal strategies and predicting the most likely outcome of a game.
A vast body of literature focuses on the computation of Nash Equilibria (NEs), mainly in two-player zero-sum games~\cite{shoham2009multiagent}.
This setting is well understood and, recently, some remarkable results have been achieved by, e.g.,~\citeauthor{brown2017safe}~\shortcite{brown2017safe,brown2017superhuman}.
While relevant, this model is rather restrictive, as many practical scenarios are not zero-sum and involve more than two players, and it presents some weaknesses when used as a prescriptive tool, in particular in general-sum games.
%
Indeed, when multiple NEs coexist, the model assumes the lack of communication between the players, preventing them from synchronizing their strategies.

In practical situations where some form of communication is possible, solution concepts different from that of NE are required.
The main alternative is the \emph{Correlated Equilibrium} (CE), introduced by~\citeauthor{aumann1974}~\cite{aumann1974}.
In a CE, a \emph{device} (i.e., a trusted external mediator) draws strategy profiles from a known joint probability distribution and privately communicates them to each player.
The probability distribution
induces an equilibrium if each player has no incentive to choose a different strategy from the recommended one, assuming the other players would not deviate either.
A variation on the CE is the \emph{Coarse Correlated Equilibrium} (CCE), introduced in~\cite{moulin1978}, which only prevents deviations happening before knowing the device's recommendation. 
In normal-form games, CEs and CCEs enjoy some appealing properties that make them plausible solution concepts in many practical scenarios.
Specifically, they arises from simple and natural learning dynamics~\cite{hart2000,cesa2006prediction}, and they can be computed via linear programming on any normal-form game in polynomial time (assuming the number of players is fixed).
Moreover, price-of-anarchy analyses show that coarse correlated equilibria characterizing outcomes of no-regret learning dynamics have near-optimal welfare~\cite{roughgarden2009intrinsic,hartline2015no}.
While {\em a} CE can be found in polynomial time in some classes of succinctly representable multi-player games, finding an {\em optimal} CE in these games is, in general, \textsf{NP}-hard~\cite{papadimitriou2008,jiang2015polynomial}.
A similar result also holds for the problem of finding an optimal CCE.
\citeauthor{barman}~\shortcite{barman} show that for graphical, polymatrix, congestion, and anonymous games the problem is \textsf{NP}-hard.
%

Sequential games allow for richer forms of interaction among the players than normal-form games, which lead to different forms of correlation whose general understanding is still limited.
Most of the works in this area focus on specific classes of games, such as Bayesian games~\cite{forges1993,forges2006} and multi-stage games~\cite{myerson1986,forges1986}.
In these specific settings, the main solution concepts studied in the literature are the \emph{Normal-Form Correlated Equilibrium} (NFCE), the \emph{Agent-Form Correlated Equilibrium} (AFCE), and the \emph{Communication-Equilibrium}.
The first two equilibria only allow for a \emph{unidirectional} communication from the device to the players, while the third equilibrium
allows for \emph{bidirectional} communication.
The only known results for general extensive-form games are due to~\citeauthor{vonStengel2008}~\shortcite{vonStengel2008}, who propose the notion of \emph{Extensive-Form Correlated Equilibrium} (EFCE).
The complex structure of extensive-form games significantly increases the computational effort required for correlation, as finding an optimal NFCE is $\mathsf{NP}$-hard even with two players~\cite{vonStengel2008}.
An optimal EFCE can be found efficiently in two-player games without Chance moves but, in games with three or more players (including Chance), finding an EFCE (or an AFCE) is $\mathsf{NP}$-hard~\cite{vonStengel2008}.
The only positive result for multi-player games is a polynomial-time algorithm to find \emph{an} EFCE~\cite{huang2008}. 

Correlated equilibria in which recommendations are drawn before the game starts are known as \emph{ex ante} CEs. These equilibria require only unilateral communication from the device to the players.
NFCE, AFCE, and EFCE belong to this family and differ in the time at which the recommendations are communicated to players.
Specifically, the NFCE requires, for each player, a single interaction with the mediator taking place before the beginning of the game, whereas AFCE and EFCE require a message for each information set reached during the game.
As a consequence, AFCE and EFCE are not suited for problems where the agents have limited communication capabilities, a situation which is frequent in practice. This is the case, for instance, of collusion in bidding, where communication during the auction is illegal, and coordinated swindling in public (see also the recent work by~\citeauthor{farina2018exante}~\cite{farina2018exante}). Different forms of correlation have been explored when a team of players faces an adversary~\cite{basilico2016,basilico2017computing,basilico2017coordinating,celli2018,farina2018exante}. This setting, also known as \emph{ex ante coordination}, is quite different from ours. Our notion of correlation is more flexible as any player may have different objectives. Therefore, in our correlation setting, individual players have to be incentivized to follow the recommendations of the mediator. In contrast, in the \emph{ex ante coordination} setting there is no need for incentive constraints since team members share their final rewards.

\subsection{Original Contributions}
In this paper, we focus on equilibria requiring a low level of communication.
A natural question is whether correlation can be reached efficiently when agents have limited communication capabilities, i.e., when they cannot receive messages during the execution of the game.\footnote{This rules out the possibility of employing an EFCE.}
Motivated by the hardness result for the NFCE, we introduce the notion of \emph{Normal-Form Coarse Correlated Equilibrium} (NFCCE) as the extension of CCE to sequential games. 

We prove that, unlike the NFCE, the problem of finding an optimal NFCCE
admits a polynomial-time algorithm for two-player games without Chance moves.
In particular, we devise a hybrid formulation (combining the normal and the sequence forms) for the problem of computing an optimal NFCCE featuring a polynomial number of constraints and an exponential number of variables.
We then provide a polynomial-time separation oracle which, thanks to the ellipsoid algorithm~\cite{khachiyan1980}, allows us to show that an optimal NFCCE can be computed in polynomial time.
We also show that this approach cannot be extended to more general settings, illustrating that with more than two players, including Chance, the problem becomes \textsf{NP}-hard.

We describe a practical algorithm to compute an optimal NFCCE based on column generation---a variation of the \emph{simplex method} in which the variables (columns) of the problem are introduced one at a time.
We devise different oracles to solve the corresponding pricing problem.
In particular, we provide a polynomial-time oracle suitable for the two-player setting, and an oracle based on a Mixed Integer Linear Program (MILP).
Then, we show how to adapt the MILP oracle to the case of two-player games with Nature, and to general multi-player games.

\section{Preliminaries}	
We briefly introduce several of the basic concepts we use in the rest of the paper.
Further details can be found in~\cite{shoham2009multiagent}.

\subsection{Game Representations}
An extensive-form game $\Gamma$ has a finite set of players $N$ and a finite set of actions $A$.
Exogenous stochasticity is represented through a non-strategic player $c$ (the nature or chance player).
$V$ is the set of non-terminal decision nodes, and $V_i\subseteq V$ is the set of decision nodes belonging to player $i\in N\cup\{c\}$.
The set of terminal nodes (leaves) is denoted by $L$. 
The function $\iota: V\to N\cup\{c\}$ associates each decision node with the player acting at it. 
The function $\rho:V\to 2^{A}$ is the \emph{action function}, assigning with each decision node a set of available actions. 
The successor function is denoted by $\chi: V\times A\to V\cup L$.
Let $U_i:L\to \mathbb{R}$ be the utility function of each each $i\in N$.
Moreover, let $U=\{U_i\}_{i\in N}$.
Finally,  for each $i\in N\cup\{c\}$ let $H_i$ be an information partition of $V_i$ such that decision nodes within the same information set $h\in H_i$ are not distinguishable by player $i$.
We write $H=\{H_i\}_{i\in N\cup\{c\}}$.
The a function  $\pi_c$ is such that $\pi_c(h,a)$ is the fixed probability with which chance selects $a$ at $h\in H_c$.
Moreover, $\rho(h)$ denotes the set of actions available at $h\in H_i$.
We remark that, by definition, $\rho(x_1)=\rho(x_2)=\rho(h)$ for any player $i\in N\cup\{c\}$, information set $h\in H_i$, and $x_1,x_2\in h$. 
In this paper, we focus on games with \emph{perfect recall}, i.e., games where, at each stage, all the players recall all the information acquired at earlier stages.

An extensive-form game can be equivalently represented in \emph{normal-form}.
Let $P_i=\times_{h\in H_i}\rho(h)$ be the set of pure normal-form plans of player $i \in N$.
A normal-form plan $p\in P_i$ specifies an action per information set of player $i$.
The \emph{normal-form} of an extensive-form game is characterized by the same set of players $N$, actions $P=\times_{i\in N}P_i$, and the set of utility functions $U' = \{U_i'\}_{i\in N}$.
Function $U_i':P\to \mathbb{R}$ denotes the expected payoff obtained by marginalizing with respect to $\pi_c$.
The \emph{reduced} normal form is obtained by deleting duplicated strategies from the normal form.

\textbf{Strategy Representations}.
A normal-form strategy $\sigma_i$ for $i\in N$ is defined as the function $\sigma_i:P_i\to \Delta^{|P_i|}$.
We denote by $\Sigma_i$ the normal-form strategy space of player $i$.
A \emph{correlated} (joint) normal-form strategy $\sigma\in\Sigma$ is defined as $\sigma: P\to\Delta^{|P|}$.
The size of a normal-form strategy is exponential in the size of the extensive-form tree.
This shortcoming can be overcome by exploiting the \emph{sequence form}~\cite{vonStengel1996}, whose size is linear in the size of the game tree.

The sequence form decomposes strategies into sequences of actions and their realization probabilities.
A {\em sequence} for player $i$, associated with a node $x$ of the game, is the subset of $A$ specifying player $i$'s actions on the path from the root to $x$.
We denote the set of sequences of player~$i$ by~$Q_i$.
A sequence is said \textit{terminal} if it leads to a terminal node for at least a set of sequences of the other players.
The set of terminal sequences of player~$i$ is denoted by~$\overline{Q}_i$.
Moreover, we denote by~$q_\emptyset$ the fictitious sequence leading to the root node and, for each action $a\in A$ and sequence $q\in Q_i$, we denote by $qa\in Q_i$ the \textit{extended} sequence obtained by appending action $a$ to $q$.

A sequence-form strategy, said \emph{realization plan}, is a function $r_i:Q_i\to \mathbb{R}$ associating each sequence $q\in Q_i$ with its probability of being played.
A well-defined sequence-form strategy is such that $r_i(q_\emptyset)=1$ for each $i\in N$ and, for each $h$ and sequence $q$ leading to $h$, $-r_{i}(q)+\sum_{a\in\rho(h)} r_{i}(qa)=0$ and $r_i(q)\geq 0$.
These constraints are linear in the number of sequences and can be compactly written as $F_i \, r_i=f_i$, where $F_i$ is an $|H_i| \times |Q_i|$ matrix and $f_i^\mathsf{T}=(1,0,\ldots,0)$ is a vector of dimension $|H_i|$.
The utility function of player $i$ is represented by a sparse $n$-dimensional matrix defined only for profiles of terminal sequences leading to a leaf node.
With a slight abuse of notation, we denote it by $U_i \in \mathbb{R}^{|Q_1| \times \dots \times |Q_n|}$.

\subsection{Correlation in Normal-Form Games}
Let $p_{-i}=(p_1,\ldots,p_{i-1},$ $p_{i+i},\ldots,p_n)\in \times_{j\in N\setminus\{i\}}P_j$.
The classical notion of CE~\cite{aumann1974} for normal-form games is:
\begin{definition}\label{def:ce}
	$\sigma^\ast\in\Sigma$ is a correlated equilibrium of the normal form game $(N,P,U')$ if,
for every $i\in N$ and $p_i,p_i'\in P_i$, the following holds:
	$$\sum\limits_{p_{-i}\in P_{-i}}\sigma^\ast(p_i,p_{-i})\left(U'_i(p_i,p_{-i})-U'_i(p_i',p_{-i})\right)\geq 0.$$
\end{definition}
\noindent A CE can be interpreted in terms of a mediator who, \emph{ex ante} the play, draws $(p_1,\ldots,p_n)$ according to the publicly known $\sigma^\ast$ and privately communicates each \emph{recommendation} $p_i$ to the corresponding player. 

Another possibility is enforcing protection against deviations of players which are independent from the sampled outcome.
This can be done though the notion of coarse correlated equilibrium~\cite{moulin1978}.
\begin{definition}\label{def:cce}
	$\sigma^\ast\in\Sigma$ is a coarse correlated equilibrium of a normal-form game $(N,P,U')$ if,
for every $i\in N$ and $p_i'\in P_i$, the following holds:
	$$\sum\limits_{p_i\in P_i}\sum\limits_{p_{-i}\in P_{-i}}\sigma^\ast(p_i,p_{-i})\left(U'_i(p_i,p_{-i})-U'_i(p_i',p_{-i})\right)\geq 0.$$
\end{definition}
CCEs differ from CEs in that a CCE only requires that following the suggested action is a best response in expectation before the recommended action is actually revealed.
Moreover, we recall that every CE is also a CCE while the converse is, in general, not true.

An optimal CCE may lead to a social welfare arbitrarily larger than the social welfare provided by the optimal CE on the same game. Figure~\ref{fig:example} reports a normal-form game where this happens ($k > 1$).

\renewcommand{\arraystretch}{2}
\begin{figure}[h!]
	\centering
	\begin{tabular}{r|c|c|c|}
		\multicolumn{1}{r}{}
		&  \multicolumn{1}{c}{$a_2^1$}
		& \multicolumn{1}{c}{$a_2^2$}
		& \multicolumn{1}{c}{$a_2^3$}\\
		\cline{2-4}
		$a_1^1$ & $k,0$ & $-k^2,0$ & $-k^2,1$ \\
		\cline{2-4}
		$a_1^2$ & $-k^2,0$ & $1,0$ & $-k^2,-1$\\
		\cline{2-4}
	\end{tabular}
	\caption{Example on the difference between CE and CCE.}
	\label{fig:example}
\end{figure}
\renewcommand{\arraystretch}{1}

The joint strategy profile assigning probability $1/2$ to $(a_1^1,a_2^1)$ and $(a_1^2,a_2^2)$ is the CCE maximizing the social welfare of the players, which is $(k+1)/2$. The unique optimal CE is the probability distribution assigning probability 1 to $(a_1^2,a_2^2)$, providing a social welfare of 1 independently of $k$. Therefore, for increasing values of $k$, an optimal CCE allows the players to reach a social welfare which is arbitrarily larger than the social welfare reached through the optimal CE.


\subsection{Correlation in General Extensive-Form Games} We review the main notions of correlation for general extensive-form games.
In this general setting, it is customary to consider \emph{ex ante} CEs, i.e., correlated equilibria in which an action profile is sampled before the game is played.
In this paper, we focus on the following solution concepts:
\begin{definition}\label{def:nfcce}
	A normal-form correlated equilibrium (normal-form coarse correlated equilibrium) of an extensive-form game $\Gamma$ is a correlated equilibrium (coarse correlated equilibrium) of the reduced normal-form game equivalent to $\Gamma$.
\end{definition}
In these two solution concepts, the entire vector of recommendations specifying one action per information set is revealed to the players before the game starts.
Thus, once the recommendation is received each player commits to playing a pure strategy.

Informally, an AFCE~\cite{forges1993} is a CE of the agent-form game equivalent to the given extensive-form game.
In the agent form of the game, moves are chosen by a different agent per information set of the player. 
In an EFCE~\cite{vonStengel2008}, each recommendation is assumed to be in a \emph{sealed envelope} and is revealed only when the player reaches the relevant information set (i.e., the information set where she can make that move). 
The main difference between EFCE and NFCE/NFCCE is that the former requires recommendations to be delivered during the game execution, thus being more demanding in terms of communication requirements.
It is crucial to notice that the size of the signal that has to be sampled is the same, and it has polynomial size (one action per information set).

Letting $S_{\circ}$ be the set of equilibria of type $\circ$ of a given game, we have: $S_{NFCE}\subseteq S_{EFCE}\subseteq S_{NFCCE}\subseteq S_{AFCE}$.
See~\citeauthor{vonStengel2008}~\shortcite{vonStengel2008} for further details.

In the next section, we study the problems of computing an NFCE and an NFCCE maximizing the social welfare (i.e., the cumulative utility of the players).
We refer to them as \textsf{NFCE-SW} and \textsf{NFCCE-SW}.
The generalization of our results to the case in which one searches for an equilibrium maximizing a linear combination of the players' utility, omitted here for reasons of space, is straightforward.

\section{Complexity of an Optimal NFCCE}

We show that there exists a polynomial-time algorithm for solving the \textsf{NFCCE-SW} problem with two players.
First, we provide a compact formulation for the problem.
Then, we describe a polynomial-time algorithm for solving it.

\subsection{Problem Formulation}
Given an extensive-form game $\Gamma$, a direct application of Definition~\ref{def:nfcce} yields a Linear Programming problem (LP) with an exponential number of variables and an exponential number of constraints.
We provide the following result:
\begin{lemma}\label{lemma:compact_formulation}
	The \textsf{NFCCE-SW} problem for an extensive-form game $\Gamma$ can be formulated as an LP with an exponential number of variables but only a polynomial number of constraints.
\end{lemma}
\noindent To prove the lemma, we provide a hybrid representation which exploits the tree structure of the problem combining both the normal form and the sequence form.
Let $r_{p_i} \in \{0,1\}^{|Q_i|}$ be a $|Q_i|$-dimensional column vector representing the pure realization plan for player~$i \in N$ that is {\em realization equivalent} to $p_i\in P_i$.\footnote{A realization plan is {\em realization equivalent} to a normal-form plan if,
for any strategy profile of the other players, they enforce the same probability distribution over the terminal nodes of the game tree.}
We recall that every plan of the reduced normal form is realization equivalent to exactly one pure realization plan, see~\citeauthor{vonStengel1996}~\shortcite{vonStengel1996}. In the following and when not differently specified, $U_i$ denotes the sequence-form utility matrix of player~$i$.

%
%
According to Definition~\ref{def:cce}, the constraints describing an NFCCE for Player~$1$ can be written as follows (for Player~$2$, the constraints are analogous):
\begin{multline*}
\sum\limits_{p_1\in P_1}\sum\limits_{p_{2}\in P_{2}}\sigma(p_1,p_2) \,U'_1(p_1,p_{2})-\\
\sum\limits_{p_1\in P_1}\sum\limits_{p_{2}\in P_{2}}\sigma(p_1,p_{2})\,U'_1(p_1',p_{2})\geq 0\qquad \forall p_1'\in P_1.
\end{multline*}

\noindent The first term is the expected utility of Player~$1$ at the equilibrium. Let $v_1$ be the $|H_1|$-dimensional vector of variables of the dual of the best-response problem in sequence form.
By definition of sequence form, $f_1^\mathsf{T}\,v_1$ is equal to the first component of $v_1$, whose value corresponds to the utility of Player~$1$ at the equilibrium.
Then:
$$
\begin{cases}
\sum\limits_{p_1\in P_1}\sum\limits_{p_{2}\in P_{2}}\sigma(p_1,p_{2}) \,U'_1(p_1,p_{2}) = f_1^\mathsf{T}\,v_1	\\
f_1^\mathsf{T}\,v_1 - \sum\limits_{p_1\in P_1}\sum\limits_{p_{2}\in P_{2}}\sigma(p_1,p_{2})\,U'_1(p_1',p_{2})\geq 0\quad \forall p_1'\in P_1
\end{cases}.
$$
\noindent The second term of the above inequalities can be written as $$\sum_{p_{2}\in P_{2}}\left(\sum_{p_1\in P_1}\sigma(p_1,p_{2})\right)\,U'_1(p_1',p_{2}).$$
Letting $\bar \sigma_2(p_2)=\sum_{p_1\in P_1}\sigma(p_1,p_2)$, $\bar \sigma_2\in\Delta^{|P_2|}$ can be interpreted as the prior probability with which plan~$p_{2}$ is played by Player~$2$. 
$\bar \sigma_2$ can be written as the following realization-equivalent sequence-form strategy: 
$\bar r_2=\sum_{p_{2}\in P_{2}}\bar \sigma(p_2)r_{p_{2}}$, which is a valid realization plan due to convexity.
Now, we only need to show that $f_1^\mathsf{T}\,v_1$ is not strictly smaller than the value of the best response of Player~$1$ given the strategy $\bar r_2$ of Player~$2$.
By exploiting the dual of the best-response problem in sequence form, this is equivalent to showing $F_1^\mathsf{T}v_1  -U_1\,\bar r_2\geq0$.
Thus, expanding $\bar r_2$ and deriving the equilibrium constraints for Player~$2$ we obtain the following mathematical program:
\begin{align}
&\max_{\sigma \geq 0,v_1,v_2}\sum_{(p_1,p_2)\in P_1\times P_2}\sigma(p_1,p_2)\,r_{p_1}^\mathsf{T}\,(U_1+U_2)\,r_{p_2}\\
&\sum_{(p_1,p_2)\in P_1\times P_2}\sigma(p_1,p_2)\,r_{p_1}^\mathsf{T}\,U_i\, r_{p_2}=f_i^\mathsf{T}v_i \quad\forall i\in N \label{eq:root}\\
&F_1^\mathsf{T}v_1 -U_1 \bigg( \sum_{p_2\in P_2} \bigg(\sum_{p_1 \in P_1}\sigma(p_1,p_2)\bigg)r_{p_2}\bigg) \geq 0 \label{eq:br_constr_1}\\
&F_2^\mathsf{T}v_2 -U_2^{\mathsf{T}}\,\bigg(\sum_{p_1\in P_1}\bigg(\sum_{p_2 \in P_2}\sigma(p_1,p_2)\bigg)r_{p_1}\bigg)\geq 0 \label{eq:br_constr_2}\\
&\sum_{(p_1,p_2)\in P_1\times P_2}\sigma(p_1,p_2)=1 \label{eq:sigma_sum_to_1}.
\end{align}

\noindent This formulation constitutes a proof of Lemma~\ref{lemma:compact_formulation} as it employs a polynomial number of constraints (namely, $|Q_1|+|Q_2|+3$) and an exponential number of variables.

\subsection{Efficient Algorithm}
The following lemma will be employed to prove our central result.
It shows that a player can reason in a \emph{best-response fashion} to minimize the utility of the other player weighted by an arbitrary distribution, while also guaranteeing the reachability of a given terminal node.
\begin{lemma}\label{lemma:backward}
	Given a generic two-player extensive-form game $\Gamma$, an outcome $\ell\in L$, and a vector $\zeta \in \mathbb{R}^{|Q_1|}$, the problem of finding $p_{2}\in P_{2}$ under the constraints that
	\begin{itemize}[nolistsep,itemsep=1mm]
		\item there exists some $p_1\in P_1$ s.t. $(p_1,p_{2})$ leads to outcome $\ell$ and
		\item $\zeta^{\mathsf{T}}\,U_1 \, r_{p_{2}}$ is minimized
	\end{itemize}
	can be solved in polynomial time. 
	The same holds when the two players are interchanged.
\end{lemma}

\noindent\textbf{Proof}.
Let us focus on the case in which we look for $p_2\in P_2$. First, define $\bar{U}_1$ s.t. $\bar{U}_1(q_1,q_2):=\zeta(q_1)U_1(q_1,q_2)$ for each $(q_1,q_2)\in Q_1\times Q_2$. Then, let $\bar{\Gamma}$ be the extensive-form game obtained from $\Gamma$ by substituting Player 1's utility function with $\bar{U}_1$. Given $\bar{\Gamma}$, denote by $(q_1^\ell,q_2^\ell)$ the pair of sequences identifying $\ell$, and by $H_i^\ell$ the set of information sets of player $i$ encountered in sequence $q_i^\ell$. Algorithm~\ref{alg:backward} returns the set of actions ($A'_i$) forming a plan of the normal-form game (not reduced) equivalent to $\bar{\Gamma}$. 

\begin{algorithm}
	\caption{\texttt{Constrained-plan-search}}
	\begin{scriptsize}
		\begin{algorithmic}[1]
			\Function{C-PLAN-SEARCH}{$x$, $\bar{\Gamma}$, $i$, $q_i^\ell$, $H_i^\ell$, $A'_i$}\Comment{$i$ is the player for which we want to find a plan, $A'_i$ is the temporary set (initially empty) of actions of $i$ selected}
			\State $\nu \leftarrow K$\Comment{$K$ is a sufficiently large constant}
			\State $a'\leftarrow$null
			\If{$x$ is terminal}
			\State \textbf{return} $(\bar{U}_{-i}(x),A'_i)$
			\Else
			\If{$x\in V_{-i}$}
			\For {$y\in x.\textnormal{child}$}
			\State $\nu +=\textnormal{C-PLAN-SEARCH}(y,\bar{\Gamma},i,q_i^\ell,H_i^\ell,A'_p).\textnormal{val}$
			\EndFor
			\State \textbf{return} $(\nu,A'_i)$
			\Else 
			\If{$\exists h \in H_i^\ell: x \in h$}
			\State $a'\leftarrow$ action specified by $q_i^\ell$
			\State $y_{a'}\leftarrow$ child of $x$ reached through $a'$
			\State $\nu=\textnormal{C-PLAN-SEARCH}(y_{a'},\Gamma',i,q_i^\ell,H_i^\ell,A'_i).\textnormal{val}$
			\Else
			\For{$y\in x.\textnormal{child}$} 							\State$temp\leftarrow\textnormal{C-PLAN-SEARCH}(y,\bar{\Gamma},i,q_i^\ell,H_i^\ell,A'_i)$
			\If{$temp.\textnormal{val}<\nu$}
			\State $\nu\leftarrow temp.\textnormal{val}$
			\State $a'\leftarrow a\in\rho(x): \chi(x,a)=y$
			\EndIf
			\EndFor
			\EndIf
			\State \textbf{return}$(\nu,A'_i\cup \{a'\})$
			\EndIf
			\EndIf
			\EndFunction
		\end{algorithmic}
	\end{scriptsize}
	\label{alg:backward}
\end{algorithm}
To retrieve $A'_i$, Algorithm~\ref{alg:backward} performs a depth-first traversal of the tree while keeping track of the value to be minimized at each decision node ($\nu$) and selecting actions while moving backwards. Then, $p_2$ can be computed by traversing the tree from the root, and selecting actions according to those specified in $A'_2$.
\hfill$\Box$

Let us focus on the dual $\mathcal{D}$ of LP~(1)--(6):
\begin{lemma}\label{lemma:sep}
	$\mathcal{D}$ admits a polynomial-time separation oracle.
\end{lemma}
\noindent\textbf{Proof}.
Let $\alpha_i\in \mathbb{R}$, for all $i \in N$, be the dual variables of constraints~(2), $\beta_{1} \in \mathbb{R}^{|Q_1|}$ the dual variables of constraints~(3), $\beta_{2} \in \mathbb{R}^{|Q_2|}$ the dual variables of constraints~(4), and $\gamma \in \mathbb{R}$ the dual variable of constraint~(5). With $n=2$, $\mathcal{D}$ is an LP with a number of variables ($|Q_1|+|Q_2|+3$) polynomial in the size of the tree and an exponential ($|P_1\times P_2| + |H_1| + |H_2|$) number of constraints. We show that, given a vector $\bar z = (\bar \alpha_1, \bar \alpha_2,\bar \beta_1,\bar \beta_2,\bar \gamma)$, the problem of either finding a hyperplane separating $\bar z$ from the set of feasible solutions to $\mathcal{D}$ or proving that no such hyperplane exists can be solved in polynomial time. Since the number of dual constraints corresponding to the primal variables $v_i$ is linear, these constraints can be checked efficiently for violation.
We are left with the problem of determining whether any of the following constraints, defined for all $(p_1,p_2)\in P_1\times P_2$, is violated:
\begin{multline*}
r^\mathsf{T}_{p_1}\,U_1\,r_{p_2} \bar\alpha_1 + r^\mathsf{T}_{p_1}\,U_2\,r_{p_2} \bar\alpha_2 + \bar\beta_1^\mathsf{T}\, U_1\, r_{p_2} +r_{p_1}^{\mathsf{T}}\, U_2\,\bar\beta_2+ \bar\gamma \geq\\
r_{p_1}^\mathsf{T}\,(U_1+U_2)\,r_{p_2}.
\end{multline*}
\noindent Let us consider the {\em separation problem} of finding an inequality of $\mathcal{D}$ which is maximally violated at $\bar z$.
The problem reads:
\begin{multline*}
\hspace{-0.6cm}
\min_{(p_1,p_2) \in P_1 \times P_2}  \Bigg\{r^\mathsf{T}_{p_1}\left((\bar \alpha_1-1) \,U_1 + (\bar \alpha_2-1) \,U_2\right) \,r_{p_2} + \\+ \bar\beta_{1}^\mathsf{T}\,U_1 \,r_{p_2}  + r_{p_1}^{\mathsf{T}}\,U_2 \,\bar \beta_{2}\Bigg\}.
\end{multline*}
\noindent A pair $p_1,p_2$ yielding a violated inequality exists iff the separation problem admits an optimal solution of value $<- \bar \gamma$.

One such pair (if any) can be found in polynomial time by enumerating over the (polynomially many) possible outcomes $\ell \in L$ of the game. For each of them, we look for the pair $(p_1^\ell,p_2^\ell)$ minimizing the objective function of the separation problem, halting as soon as a pair $(p_1',p_2')$ yielding a violated constraint is found. If the procedure terminates without finding any suitable pair, we deduce that no violated inequalities exist and $\mathcal{D}$ has been solved. 
First, notice that $r^\mathsf{T}_{p_1}((\bar \alpha_1-1) U_1 + (\bar \alpha_2-1) U_2) r_{p_2}$ is constant for the family of pairs identifying $\ell\in L$. Therefore, we can consider an individual subproblem for each player (i.e., we can find $p_1^\ell$ and $p_2^\ell$ independently).
Hence, for each outcome $\ell$ and for each player $i$ the corresponding $p_i^\ell$ can be found in polynomial time due to Lemma~\ref{lemma:backward}.
\hfill$\Box$

The following theorem shows that, in certain cases, the \textsf{NFCCE-SW} problem can be solved efficiently:
\begin{theorem}\label{th:nfcce_poly}
	Given an extensive-form game $\Gamma$ with $n=2$ players and without chance moves, an NFCCE maximizing the social welfare can be computed in time polynomial in the size of the game tree.
\end{theorem}
\noindent\textbf{Proof}. Lemma~\ref{lemma:sep} shows that there exists a polynomial-time separation oracle for $\mathcal{D}$.
Then, $\mathcal{D}$ can be solved in polynomial time via the ellipsoid method due to the equivalence between optimization and separation~\cite{khachiyan1980,Grotschel1981}.
%
As the method solves, in polynomial time, a primal-dual system encompassing not just $\mathcal{D}$ but also its primal problem~\textsf{NFCEE-SW}, it also produces, simultaneously,  an optimal solution to the latter.\hfill$\Box$


\subsection{Negative Result}
%
The approach that we presented here cannot be extended to games with two players and the chance player as, upon introducing the latter, the problem transitions from polynomially solvabile to \textsf{NP}-hard:\footnote{Other problems in which this transition takes place are, for example, the problem of computing a socially optimal EFCE~\cite{vonStengel2008} and the problem of deciding if a two-player zero-sum extensive-form game with perfect recall admits a pure strategy equilibrium~\cite{blair1996perfect,hansen2007finding}.}
%
%
\begin{theorem}\label{th:hard}
	Computing an NFCCE maximizing the social welfare is \textsf{NP}-hard even in extensive-form games with two players, chance moves, and binary outcomes.
\end{theorem}
\textbf{Proof Sketch}.
	A construction introduced by~\citeauthor{vonStengel2008}~\cite{vonStengel2008} can be employed.  The reduction is from \textsf{SAT}, whose generic instance is a Boolean formula $\phi$ in conjunctive normal form with $\eta$ clauses and $\nu$ variables. Given $\phi$, we build an auxiliary game $\Gamma_\phi$, of size proportional to that of the boolean formula, following~\cite[Theorem 1.3]{vonStengel2008}.
	$\Gamma_\phi$ admits a pure strategy guaranteeing a social welfare of 2 if and only if $\phi$ is satisfiable. Otherwise, the maximum expected social welfare cannot be more than $2(1-1/\eta)$. A pure strategy maximizing the social welfare is also an NFCCE, since no \emph{ex 
	ante} deviation would result in an increase in the player's utility, being it already maximal. Then, finding a solution to \textsf{NFCCE-SW} in polynomial time would imply the existence of a polynomial time algorithm for \textsf{SAT}, which leads to a contradiction, unless \textsf{P}=\textsf{NP}.
\hfill$\Box$

\noindent Notice that, when considering the separation problem of $\mathcal{D}$, working with chance is hard because the first term of the objective function of the separation problem is no longer constant when the outcome is fixed.
In the case with $n>2$ and no chance moves, one would have to determine the joint best response of two player a time (to maximize the terms of the objective function of the separation problem following the first one), which is \textsf{NP}-hard~\cite{vonStengel2008}.

\section{A Practical Algorithm}\label{sec:practical_alg}

%
Due to being based on the ellipsoid method (which, while being a powerful theoretical tool, is well-known to be inefficient in practice), the algorithm that we used in the proof of Theorem~\ref{th:nfcce_poly} is not appealing from a practical perspective.
%
We propose, here, a computationally more efficient method based on the simplex method
to compute optimal NFCCEs via a {\em column generation} technique.
The focus on two-player games is motivated by the negative result in the previous section.

Let $x$ be a vector containing the variables of LP~(1)--(6):
$$x^\mathsf{T}=(\underbrace{\sigma(p_1',p_2'),
\ldots,\sigma(p_1'',p_2'),\ldots}_{|P_1\times P_2|},v^\mathsf{T}_1,v^\mathsf{T}_2,s^\mathsf{T}_1,s^\mathsf{T}_2),$$

\noindent where, for each $i = 1,2$, $v_i$ is defined as in the proof of Lemma~\ref{lemma:compact_formulation} and $s_i$ is a $|Q_i|$-dimensional column vector of slack variables.
The cost vector $c$ associated with the variables is:
%
$$c^\mathsf{T}=(\underbrace{[U_1'(p_1',p_2')+U_2'(p_1',p_2')]}_{\sigma(p_1',p_2')},\ldots,\underbrace{0,\ldots,0}_{|H_1|+|H_2|+|Q_1|+|Q_2|}),$$ 
\noindent where $U_i'$ is the utility matrix of the reduced normal-form game.
We compactly rewrite the constraints of LP~(1)--(6) in standard form as $M\,x=b$, where $b^\mathsf{T}=(1,0,\ldots,0)$ is a vector of dimension $(|Q_1|+|Q_2|+3)$.
%
%
%
We denote the $j$-th column of $M$ by $M_{(\cdot,j)}$.

The algorithm works in two phases, determining, first, a basic feasible solution and, then, iteratively improving it until an optimal one is found.
%
%
The crucial component of the algorithm is an oracle for solving, given a basic feasible solution to LP (1)--(6), the problem (we refer to it as \textsf{LRC}) of finding a variable with the largest reduced cost.
%
%
Notice that Theorem~\ref{th:nfcce_poly} already implies the tractability of the problem of finding the variable with the maximum reduced cost---the so-called (primal) \emph{pricing} problem, as it is equivalent to finding a maximally violated constraint in the dual $\mathcal{D}$.
Hence:
\begin{corollary}\label{cor:pricing}
\textsf{LRC} can be solved in polynomial-time.   
\end{corollary}


Letting $c_j$ be the cost associated with the $j$-th component  of $x$ and letting $c_B$ be the vector of costs of the basic variables, the $j$-th reduced cost is:
\begin{equation}\label{eq:reduced_cost}
\overline{c}_j=c_j-c_B^\mathsf{T}\,B^{-1}\,M_{(\cdot,j)},
\end{equation}
where $B=[M_{(\cdot,j')},M_{(\cdot,j'')},\ldots]$ for each index $j',j'',\ldots$ corresponding to a basic variable.
%
We rely on the following polynomial-time oracle, \textsf{P-LRC}, described in Algorithm~2 (another oracle is presented in the next section).

First, notice that, given a basic feasible solution, $c_B^\mathsf{T} B^{-1}$ is equal to a vector (call it $\zeta$) of dimension $(|Q_1|+|Q_2|+3)$, computable in polynomial time (Line~4).
By employing the same notation as the one adopted for the dual variables in the proof of Lemma~\ref{lemma:sep}, $\zeta^\mathsf{T}=(\bar \beta_1,\bar \beta_2, \bar \alpha_1,\bar \alpha_2,\bar \gamma)$, where $\bar\beta_i$ is the vector of dual variables of constraints~\eqref{eq:br_constr_1} and~\eqref{eq:br_constr_2}, $\bar \alpha_i$ are the dual variables of constraints~\eqref{eq:root}, and $\bar \gamma$ is that of constraint~\eqref{eq:sigma_sum_to_1}.

\begin{algorithm}[H]
	\caption{\textsf{P-LRC}}
	\begin{footnotesize}
		\begin{algorithmic}[1]
			\Function{P-LRC}{$\Gamma$, $M$, $c$, $B$}
			\State $J\leftarrow\emptyset$
			\State $\forall j, \bar c_j\leftarrow\infty $
			\State $\zeta\leftarrow c_B^\mathsf{T}B^{-1}$\label{line:z}
			\For{$j\in\{|P_1\times P_2|+1,\ldots,|c|\}$} \label{line:compute_v_s_start}
			\State $\bar c_j\leftarrow c_j-\zeta M_{(\cdot,j)}$
			\State $J\leftarrow J\cup\{j\}$			\label{line:compute_v_s_end}	
			\EndFor
			\For{$\ell\in L$} \label{line:iterate_l}
			\State $\hat p_i\leftarrow\textnormal{\textsf{C-PLAN-SEARCH}}(\ell,\bar \beta_i)$, $\forall i\in N$ \label{line:plan_search}
			\State $\hat j\leftarrow$ index of $\sigma(\hat p_1, \hat p_2)$ in $c$
			\State $\bar c_{\hat j}\leftarrow c_{\hat j}-\zeta M_{(\cdot,\hat j)}$
			\State $J\leftarrow J\cup\{\hat j\}$\label{line:iterate_l_end}	
			\EndFor
			\State $j^\ast=\argmax_{j\in J} \bar c_j$ \label{line:argmax}
			\State \textbf{return} $j^\ast$
			\EndFunction
		\end{algorithmic}
	\end{footnotesize}
	\label{alg:plrc}
\end{algorithm}

The reduced costs of the variables $v_i$ and $s_i$ can be computed directly by definition since their number is polynomial in the size of the tree (Lines~5 to~7).
We are left with the problem of evaluating the reduced costs of the $\sigma(\cdot,\cdot)$ variables.
\textsf{P-LRC} enumerates the outcomes of the game (Line~8).
Since all the pairs of plans identifying $\ell$ have the same $c_j$, the problem of minimizing $\zeta^\mathsf{T} M_{(\cdot,j)}$ amounts to finding a pair $(p_1,p_2)$ minimizing $(\bar\beta_1 U_1r_{p_2}+\bar \beta_2 U_2^\mathsf{T}r_{p_1})$.
The problem can be split into a subproblem per player, and solved through Algorithm~\ref{alg:backward}, which we presented in the proof of Lemma~\ref{lemma:backward} (Line~9, where we simplified the signature of \textsf{C-PLAN-SEARCH} for ease of notation).
By applying this procedure for each of the outcomes and selecting, among the resulting pairs, the one with the largest reduced cost (Line~13), we are able to determine the new variable entering the basis in polynomial time.

The two phases of the overall algorithm are the following ones, and both adopt \textsf{P-LRC}:


\textbf{Phase 1: finding a feasible point}.
A basic feasible solution to \textsf{NFCCE-SW} is determined through an auxiliary problem with artificial variables, 
where a new variable is introduced for each equality constraint, and their sum is minimized in the objective function.
If some artificial variable with index $\bar j$ is found in the optimal basis of the auxiliary problem, we can find, in polynomial-time, a variable~$j$ of the original problem to replace it by either maximizing or minimizing $e_{\bar j}B^{-1}M_{(\cdot,j)}$, where $e_{\bar j}$ is a vector of zeros with suitable dimension and equal to 1 in position $j$ (the problem can be solved with Algorithm~\ref{alg:backward}). 

\textbf{Phase 2: finding an optimal solution}. 
Starting from a basic feasible solution, the algorithm iteratively improves it until an optimal solution is found.
While, if we were to solve the problem with a standard implementation of
the simplex method,
we would have to
compute the reduced cost of all the nonbasic variables to find one to enter the basis (which would require exponential time in the size of the game), by employing \textsf{P-LRC} the next variable to enter the basis can be found in polynomial time.
%
%
This follows from the same reasoning that led to Corollary~\ref{cor:pricing}.

We remark that, while the two phases require polynomial time, the bottleneck of the approach is that, at each iteration, \textsf{P-LRC} has to traverse the game tree twice for each $\ell\in L$.
%
To circumvent this issue, we present a second oracle based on mixed-integer linear programming (see the experimental evaluation for a comparison between the two approaches).

\section{General Mixed-Integer Oracle}

In this section, we describe an oracle (\textsf{MI-LRC}) for computing a solution to \textsf{LRC} by solving a Mixed-Integer Linear Program (MILP).
Differently from \textsf{P-LRC}, \textsf{MI-LRC} does not need the explicit enumeration of the terminal nodes of the game, and, furthermore, it can be extended to games with chance and more that two players.
We provide, here, a description of the oracle for the case of a two-player game with and without chance moves.\footnote{\textsf{MI-LRC} can be extended to games with $n>2$, we omit the description of this setting due to space constraints.}

The crucial difference between \textsf{MI-LRC} and \textsf{P-LRC} is in the way they handle the inspection of the reduced costs associated with the $\sigma(\cdot,\cdot)$ variables.
In \textsf{MI-LRC}, lines~8--12 of Algorithm~2 are substituted with an MILP.

\subsection{Two-player games} Let us first focus on the case of a two-player game without chance moves.
Let $R_i$ be a $|Q_i|\times |L|$ matrix such that $R_i(q_i,\ell)=1$ if $q_i$ is on the path from the root to $\ell$, and $R_i(q_i,\ell)=0$ otherwise.
Let also $z$ be an $|L|$-dimensional vector of binary variables.
\textsf{MI-LRC} solves the following problem: 
\begin{align}
	\hspace{-1cm}\max_{\substack{z \in \{0,1\}^{|L|}\\r_i \in \mathbb{R}^n_+}} & \left((1 - \bar\alpha_1)r_1^\mathsf{T} - \bar \beta_1^\mathsf{T}\right) U_1 r_2 + r_1^\mathsf{T} U_2\left((1 - \bar\alpha_2)r_2  - \bar\beta_2\right) \label{eq:obj}\\
	&F_i r_i=f_i\qquad \forall i\in N\\
	&r_i\geq R_{i}z\qquad \forall i\in N \label{eq:R}\\
	&\sum_{\ell\in L}z(\ell)=1\label{eq:sum_to_one}.
\end{align}
The objective function~\eqref{eq:obj} follows from the definition of the reduced costs (we are looking for a variable whose dual constraint is maximally violated).
Constraints~\eqref{eq:R} force the realization plans to select with probability 1 the sequences on the path to the selected outcome $\ell$.
Notice that, while the objective function contains quadratic terms, they only involve binary variables.
Therefore, it can be restated as a linear function after introducing a new variable and four linear constraints per bilinear term according to the formulation proposed in~\cite{mccormick1976computability}.

Notice that an optimal realization plan $r^\ast_i$, solution to \textsf{MI-LRC}, may not be \emph{pure} (i.e., there may exist some $q\in Q_i$ s.t. $r^\ast_i(q)\in(0,1)$). Nevertheless, there always exists a pair of pure realization plans leading to the same terminal node and granting the same value $\bar{\beta_1^\mathsf{T}}U_1r^\ast_2+r^{\ast\mathsf{T}}_1U_2\bar{\beta_2}$.
%
%
Once a pair of pure realization plans has been determined, the reduced cost associated with it has to be computed according to equation~\eqref{eq:reduced_cost} and compared to the reduced costs of the remaining variables (Line~13 of Algorithm~2).\footnote{It is enough to traverse the tree depth-first, and select sequences, among those played with strictly positive probability in $r_i^\ast$, following the same reasoning of Algorithm~\ref{alg:backward}.}

\subsection{Two-player games with Nature}
We denote by $(q_1^\ell,q_2^\ell,q_c^\ell)$ the unique tuple of the sequences leading to $\ell$, where $q_c^\ell$ is a sequence of the chance player.
The crucial point is that, given $\ell\in L$, there may exist some $\ell'\in L\setminus\{\ell\}$, reachable through $(q_1^\ell,q_2^\ell,q_c^{\ell'})$, satisfying $q_c^{\ell'}\ne q_c^\ell$.
\textsf{MI-LRC} can be adapted to this scenario as follows.
First, for each $i\in N$ we compute the utility matrices $U_{i,\pi_c}$ (with dimension $|Q_1|\times|Q_2|$) obtained by marginalizing each $U_i$ with respect to $\pi_c$.
Formally, denoting by $r_c$ the realization plan defined over the sequences of the chance player which are realization-equivalent to $\pi_c$, for each $(q_1,q_2)\in Q_1\times Q_2$ we have $U_{i,\pi_c}(q_1,q_2)=\sum_{q_c\in Q_c} r_c(q_c) U_i(q_1,q_2,q_c)$.
Objective function~\eqref{eq:obj} is then modified by substituting each $U_i$ with $U_{i,\pi_c}$.
Moreover, upon denoting by $R_c$ the $|\bar Q_c|\times |L|$ matrix defined analogously to $R_i$, it suffices to substitute each of constraints~\eqref{eq:sum_to_one}, one per $\bar q\in \bar Q_c$, with the
%
%
constraint $R_{c,(q_c,\cdot)}z=1$,
where $R_{c,(q_c,\cdot)}$ denotes row $q_c$ of $R_c$.
This way, \textsf{MI-LRC} can be extended to the more demanding setting of games with two-players and chance moves. 

\subsection{Multi-player games}

We focus on a game with $n=3$, without chance moves. The oracle can be easily adapted to the setting with $n>3$, and to include the Chance player. Denote by $U_{i,r_j}$ the utility matrix of player $i$ marginalized with respect to realization plan $r_j$ of player $j$--- notice that the $U_{i,r_j}\in \mathbb{R}^{|Q_i|\times |Q_t|}$, with $i,t\in N$, and $i,t\ne j$. The objective function that needs to be maximized is:
\begin{equation}\label{eq:3_obj}
\begin{split}
r_1^\mathsf{T}\left( \left( 1-\bar\alpha_1\right)U_{1,r_3}+\left( 1-\bar\alpha_2\right)U_{2,r_3}+\left( 1-\bar\alpha_3\right)U_{3,r_3}\right)r_2\\
-\bar\beta_1^\tr U_{1,r_3}r_2 - \bar\beta_2^\tr U_{2,r_3}^\tr r_1 - \bar\beta_3^\tr U_{3,r_1}^\tr r_2
\end{split}
\end{equation}
The first term of the objective function only depends on the choice of a single terminal node $\ell\in L$. 
The following terms can be addressed following the same reasoning we employed to adapt \textsf{MI-LRC} to the case of a two-player game with Chance.
For example, in $-\bar\beta_1^\tr U_{1,r_3}r_2$ player 2 and 3 are jointly best-responding against a fixed distribution of player 1 ($\bar\beta_1$).
Then, \textsf{MI-LRC} can be substituted with the following oracle:
\vspace{0.7cm}
\begin{align}
\begin{split}
\max \bigg(r_1^\tr\big( \left( 1-\bar\alpha_1\right)U_{1,r_3}+\left( 1-\bar\alpha_2\right)U_{2,r_3}\\+\left( 1-\bar\alpha_3\right)U_{3,r_3}\big)r_2\\
-\bar\beta_1^\tr U_{1,r_3}r_2 - \bar\beta_2^\tr U_{2,r_3}^\tr r_1 - \bar\beta_3^\tr U_{3,r_1}^\tr r_2\bigg)
\end{split}\label{eq:3_obj}\\
&F_i r_i=f_i\qquad \forall i\in N\\
&z\in\{0,1\}^{|L|}\\
&\sum_{\ell\in L} z(\ell)=1 \label{eq:3_sum1}\\
&z_i\in\{0,1\}^{|L|}\qquad \forall i\in N \\
& R_{i,(q_i,\cdot)}z_i=1 \qquad \forall i\in N,\forall q_i\in \bar Q_i \label{eq:3_zi_R}\\
& z_i(\ell)\geq z(\ell) \qquad \forall i\in N, \forall \ell\in L \label{eq:3_zi_z}\\
& r_j\geq R_iz_i \qquad \forall i\in N,\forall j\in N\setminus\{i\} \label{eq:3_r_zi}\\
&r_i\geq 0\qquad \forall i\in N
\end{align}

\vspace{0.7cm}
\noindent The oracle employs $n+1$ $|L|$-dimensional vectors of binary variables. Vector $z$ selects a single terminal node (constraint~\ref{eq:3_sum1}), determining the value of the first term of the objective function. Each $z_i$, instead, selects the terminal nodes reachable through a certain choice of plans of the players that are best-responding against $i$ (constraint~\ref{eq:3_zi_R}). As an example, $z_1(\ell)=1$ iff $\ell$ is reachable through the chosen $(r_2,r_3)$. Realization plans are constrained to be consistent with the selected outcomes (constraint~\ref{eq:3_r_zi}). Finally, the choices in $z$ and in each $z_i$ have to be mutually consistent (constraint~\ref{eq:3_zi_z}).
\vspace{1cm}

\section{Discussion}
In this paper, we have studied {\em ex ante} correlated equilibria in extensive-form games with low communication requirements.
First, we showed that an optimal NFCCE can be computed in polynomial time in two-player games.
Moreover, we have devised a column generation method which allows for computing solutions iteratively, by employing one of the two oracles which we have devised for the problem of finding a column with the largest reduced cost.
In the future, it would be interesting experimentally evaluate our techniques, and to eventually further improve the scalability of our methods to tackle practical problems.
Among the possible techniques to achieve this, we mention the adoption of heuristics for solving our oracle, the use of stabilization techniques, and the introduction of dominance relationships among the columns.

\bibliographystyle{aaai}
\bibliography{biblio}

\end{document}